%% file: main.tex
\documentclass{article}
\usepackage[utf8]{inputenc}
\usepackage{color}
\usepackage[margin=1in,footskip=0.25in]{geometry}
\usepackage[maxbibnames=99,backend=bibtex]{biblatex}
\bibliography{WWW.bib}
\usepackage{tabularx}
\usepackage{bbold}
\usepackage{graphicx}
\usepackage{caption}
\usepackage{subcaption}
\usepackage[utf8]{inputenc} 
\usepackage[T1]{fontenc}    
\usepackage{hyperref}       
\usepackage{url}            
\usepackage{booktabs}       
\usepackage{amsfonts}       
\usepackage{nicefrac}       
\usepackage{microtype}      

\usepackage[utf8]{inputenc}
\usepackage{esvect}
\usepackage{amsmath}
\usepackage{comment}
\usepackage{enumerate}  
\usepackage{paralist}
\usepackage{graphicx}
\usepackage[utf8]{inputenc} 
\usepackage[T1]{fontenc}    
\usepackage{hyperref}       
\usepackage{url}            
\usepackage{booktabs}       
\usepackage{amsfonts}       
\usepackage{nicefrac}       
\usepackage{microtype}      
\usepackage{amsmath}
\usepackage{listings}
\usepackage{color}

\definecolor{codegreen}{rgb}{0,0.6,0}
\definecolor{codegray}{rgb}{0.5,0.5,0.5}
\definecolor{codepurple}{rgb}{0.58,0,0.82}
\definecolor{backcolour}{rgb}{0.95,0.95,0.92}
\definecolor{mygreen}{RGB}{28,172,0} 
\definecolor{mylilas}{RGB}{170,55,241}

\usepackage{algorithm}
\usepackage{algpseudocode}
\usepackage{pifont}
\usepackage{authblk}

\algblock{ParFor}{EndParFor}
\algnewcommand\algorithmicparfor{\textbf{parfor}}
\algnewcommand\algorithmicpardo{\textbf{do}}
\algnewcommand\algorithmicendparfor{\textbf{end\ parfor}}
\algrenewtext{ParFor}[1]{\algorithmicparfor\ #1\ \algorithmicpardo}
\algrenewtext{EndParFor}{\algorithmicendparfor}

\usepackage{tikz}
\usetikzlibrary{fit,positioning,calc,shapes.misc}

\DeclareMathOperator*{\argmin}{\arg\!\min}

\algnewcommand{\Inputs}[1]{%
  \State \textbf{Inputs:}
  \Statex \hspace*{\algorithmicindent}\parbox[t]{.8\linewidth}{\raggedright #1}
}
\algnewcommand{\Initialize}[1]{%
  \State \textbf{Initialize:}
  \Statex \hspace*{\algorithmicindent}\parbox[t]{.8\linewidth}{\raggedright #1}
}

\begin{document}
\title{\textbf{Social Discrete Choice Models}}
\author[2]{Danqing Zhang}
\author[1,3]{Kimon Fountoulakis}
\author[4]{Junyu Cao}
\author[1,3]{Michael W. Mahoney}
\author[2]{Alexei Pozdnoukhov}
\affil[1]{International Computer Science Institute (ICSI), University of California, Berkeley}
\affil[2]{Systems Engineering, University of California, Berkeley}
\affil[3]{Statistics, University of California, Berkeley}
\affil[4]{Industrial Engineering \& Operations Research, University of California, Berkeley}
\affil[ ]{\textit{\{danqing0703,kfount,jycao,alexeip\}@berkeley.edu\},mmahoney@stat.berkeley.edu}}                   
\setcounter{Maxaffil}{0}
\renewcommand\Affilfont{\itshape\small}
\maketitle









\begin{abstract}
Human decision making underlies data generating process in multiple application areas, and models explaining and predicting choices made by individuals are in high demand. 
Discrete choice models are widely studied in economics and computational social sciences. As digital social networking facilitates information flow and spread of influence between individuals, new advances in modeling are needed to incorporate social information into these models in addition to characteristic features affecting individual choices.

In this paper, we propose two novel models with scalable training algorithms: local logistics graph regularization (LLGR) and latent class graph regularization (LCGR) models.  
We add social regularization to represent similarity between friends, and we introduce latent classes to account for possible preference discrepancies between different social groups. Training of the LLGR model is performed using alternating direction method of multipliers (ADMM), and training of the LCGR model is performed using a specialized Monte Carlo expectation maximization (MCEM) algorithm. Scalability to large graphs is achieved by parallelizing computation in both the expectation and the maximization steps.
The LCGR model is the first latent class classification model that incorporates social relationships among individuals represented by a given graph. 

To evaluate our two models, we consider three classes of data: small synthetic data to illustrate the knobs of the method, small real data to illustrate one social science use case, and large real data to illustrate a typical large-scale use case in internet and social media applications. We experiment on synthetic datasets to empirically explain when the proposed model is better than vanilla classification models that do not exploit graph structure. We illustrate how the graph structure and labels, assigned to each node of the graph, need to satisfy certain reasonable properties. We also experiment on real-world data, including both small scale and large scale real-world datasets, to demonstrate on which types of datasets our model can be expected to outperform state-of-the-art models.
\end{abstract}



\input{Introduction}
\input{Literature_review}
\input{social_latent}
\input{learning}
\input{Experiments}

\input{conclusion}

\input{appendix}

\printbibliography
\end{document}

%% file: Introduction.tex
\section{Introduction}
In this paper, we focus on how to efficiently incorporate social network information into latent class models for user discrete choice modeling problems. Traditional models ignore social information and make the assumption that labels are separable in the feature space. However, for many life-style related choices (such as bicycling vs driving to work, smoking vs not smoking, overeating vs not overeating), social considerations are thought to be a key factor. People with very similar characteristics can have very different choices, often thought to be due to the influence of their friends. Also, people who make similar lifestyle related choices are thought to be more likely to be connected and form communities. Although these "birds of a feather flock together" phenomena are widely studied in social sciences, no existing predictive model can efficiently solve this problem in computational social science. In this paper, we reformulate the problem from discrete choice settings into a graph-based semi-supervised classification problem.

We propose two models to efficiently exploit the social network (graph) information. (1) The first model is the local logistics graph regularization (LLGR) method. Parameter estimation of this model is performed using a specialized Alternating Direction Method of Multipliers (ADMM), where the computation of each node can be parallelized, making the algorithm very scalable to large graphs. (2) The second model is the latent class graph regularization (LCGR) model, where we aim to combine the expressiveness of parametric model specifications with descriptive exploratory power of latent class models. Parameter estimation of the LCGR model is performed using a specialized Monte Carlo expectation maximization algorithm presented in Section~\ref{sec:param_est}. We adopt the same ADMM techniques for the M step and discuss the  parallel computation for the E step in Section~\ref{subsubsec:scalestep}. The LCGR model can outperform the LLGR model, but it is computationally more expensive. We recommend using the LCGR model for small graphs and the LLGR model for large graphs (both of which are of interest in web applications). 

To illustrate the usefulness of our methodology, we look at three classes of data.
(1) The first class is small synthetic data used to illustrate how the knobs of our methods perform in idealized and less-than-idealized situations. We experimented on our methods by tuning the class connectivity hyperparameter $\beta$ and choice preference hyperparameter $w$. When labels are not separable in feature space (which means linear hyperplanes that separate the data $x_i$ accurately do not exist), but are separable in the graph space (which means decisions $y_i$ are clustered based on communities in the graph), our model outperformed all other baseline models. In other cases, our model performed no worse than other models.
(2) The second class is small real data used to illustrate how our method performs on a typical example of interest to social scientists, and we compared with the state-of-arts methods in social science. We experimented our models on real world adolescent smoking dataset from 1995 to 1997. We found out that the smoking preferences are largely defined by the objective factors for those adolescents at first in 1995. But smoking within a certain group of teenagers became a social norm in 1997, and our social discrete choice model performed much better than models that ignore social networks structure. 
(3) The third class is a large-scale example from internet analysis used to illustrate how our method can be expected to perform in larger-scale internet and social media applications, compared with other scalable methods \cite{grover2016node2vec,defferrard2016convolutional,kipf2016semi}. A large scale experiment is conducted on an online retail account relation graph for fraud detection. Our method is more robust than other semi-supervised graph-based classification methods on a graph with huge components and high average degree, which is very common in real world applications.

%% file: Literature_review.tex
\section{Literature Review}
\label{sec:literature}
Social discrete choice model is a topic that both social scientists and computer scientists are interested in, yet they approach this topic in different ways. Social scientists propose various models with latent class, and most of the time focus on small graphs. Computer scientists generalize this problem to graph-based semi-supervised classification, and come up with various of methods to deal with large graphs.
\vspace{-3 mm}
\subsection{Discrete Choice Modeling and Latent Class Model}
\label{sec:literature1}
In many application domains, human decision making is modeled by discrete choice models. These models specify the probability that a person chooses a particular alternative from a given choice set, with the probability expressed as a function of observed and unobserved latent variables that relate to the attributes of the alternatives and the characteristics of the person. Multinomial logit models are in the mainstream of discrete choice models, with maximum likelihood used for parameter estimation from manually collected empirical data. It is important for practitioners to interpret the observed choice behaviors, and models that are linear in parameters are most common. At the same time, choice preferences within different social groups (though seemingly similar in terms of the observed characteristics) can vary significantly due to the unobserved factors or different context of the choice process. One way of accounting for this is to introduce latent class models. Latent class logistic regression models are common tools in multiple domains of social science research~\cite{CFA2006, RLN1999, WL2007}. 

It is also recognized that social influence can be a strong factor behind variability in choice behaviors. The impact of social influence on individual decision-making has attracted a lot of attention. Researchers have employed laboratory experiments, surveys, and studied historical datasets to evaluate the impact of social influence on individual decision making. However, it is difficult to avoid an identification problem in the analysis of influence processes in social networks~\cite{Manski1993}. One has to account for endogeneity in explanatory variables in order for claims of causality made by these experiments to be useful~\cite{Dugundji2005}. Due to these limitations of observational studies of influence, randomized controlled trials are becoming more common. In general, distinguishing social influence in decision making from homophily, which is defined as the tendency for individuals with similar characteristics and choice behaviors to form clusters in social networks, is currently a growing area of research and debate~\cite{Shalizi2011, Christakis2013}.

\subsection{Graph-based semi-supervised classification}
\label{sec:literature2}
Our graphical extension to the latent class model reformulates the social discrete choice problem into a semi-supervised graph-based classification problem, for which three mainstream solutions exist.

\subsubsection{Graph Regularization approach}
Graph regularization methods that penalize parameter differences among the connected nodes have been studied in the context of classification, clustering, and recommendation systems~\cite{ACC2010, MZLLK2011}. Graph-based semi-supervised learning of this kind adds a graph Laplacian regularization term to the objective function of supervised loss \cite{zhu2003semi,zhou2004learning,wijaya2013pidgin}. The graph 
Laplacian regularization term in the loss function is shown as below, where $\Delta$ is the graph Laplacian matrix:
\begin{equation}
\lambda \sum_{(i,j) \in \mathcal{E}} a_{i,j} ||f(x_i)-f(x_j)||^{2} = \lambda f^{T}(A-D)f = \lambda f^{T}\Delta f 
\end{equation}
These models assume that connected nodes tend to have similar probability distribution of labels, which means nearby nodes in a graph are likely to have the same labels. For parameter estimation, Zhu et al. \cite{zhu2003semi} and Zhou et al. \cite{zhou2004learning} proposed diffusion-based learning algorithms that involve solving linear systems directly using matrix operations. Gleich et al.\cite{gleich2015using} reformulated the diffusion-based learning problem into an optimization and added l1 regularization term to obtain a more robust solution. And a local push algorithm as in \cite{andersen2006local} was introduced to calculate local solution.

On the other hand, our proposed semi-supervised latent class classification has simpler problem formulation. Our model assumes that connected nodes tend to have similar local classifier. That means connected nodes have similar probability distribution of labels only when they have similar feature vectors. Our graph regularization objective function is shown below:
\begin{align}
\min \sum_{i \in \mathcal{V}}  \log \left(1+e^{-y_i \times (W^{T}x_i +b_i)}\right) + \lambda\sum_{(i,j)\in \mathcal{E}}  (b_{i}-b_{j})^2 
\end{align}
In addition, we used the ADMM method to speed up the algorithm by distributing the computation, owing to recent advances in distributed optimization applied to parametric models on networks~\cite{hallac2015network}.

\subsubsection{Random-walk based index-context pair approach}

The recently developed skip-gram model is widely used in learning word embedding \cite{mikolov2013distributed,mikolov2013efficient} and node embedding, both in unsupervised and semi-supervised manner. The objective is to maximize the probability of observing a context based on an index (node), where the context can either be k-hops neighbors of the node \cite{tang2015line} or the random path of the node \cite{perozzi2014deepwalk,grover2016node2vec}. By adding an extra supervised term to the objective function, both node embedding and classification tasks can be done simultaneously \cite{yang2016revisiting}.

\subsubsection{Deep Learning Approach}

Recent development in deep learning has extended the convolutional networks idea from image to graphs \cite{defferrard2016convolutional,henaff2015deep,kipf2016semi}. These models are feedforward neural networks that directly apply spectral convolution operations to inputs. Henff et al.~\cite{defferrard2016convolutional} used K-localized convolution to replace the spectral convolution operations, Kipdf et al.~\cite{kipf2016semi} used linear model as first-order approximation of localized spectral filters. Graph Convolutional Networks (GCN)~\cite{kipf2016semi} is computationally less expensive than CNN on graphs \cite{defferrard2016convolutional}, and the authors claimed the model can outperform all other models on the public dataset for semi-supervised classification problem.

%% file: social_latent.tex
\section{Social Models}
\label{sec:social_models}
We define $[N] := \{1,2,\dots,N\}$, $i\in[N]$, $t \in [K]$, where $N$ and $K$ are integers. We will use the following notations and definitions.\\

\begin{table}[h]
  \centering
  \caption{Notation: Table for notations}
    \begin{tabular}{|p{2cm}|p{5cm}|}
      \hline
    Variable & Definition \\
    \hline
    $N$ &  number of individuals\\
     \hline
    $K$  & number of latent classes\\
     \hline
    $n$ & number of samples per node\\
     \hline
    $d$ & number of features for each individual\\
     \hline
    $x_{i} \in \mathbb{R}^{d\times n}$ & feature-samples matrix of individual $i$\\
     \hline
    $z_{i}\in [K]$ & latent class variable of individual $i$\\
     \hline
    $y_{i} \in \{-1,1\}$ &binary choice of individual $i$\\
     \hline
    $W_{t} \in \mathbb{R}^{d}$ &model coefficients of class $t$\\
     \hline
    $b_{it} \in \mathbb{R}$& model offset coefficients of individual $i$ with class $t$\\
     \hline
    $\mathcal{V}$& set of nodes in a social graph, with each node corresponding to an individual\\
     \hline
    $\mathcal{E}$\footnote{We assume that each edge in the graph is represented in the set $\mathcal{E}$ only once.}& set of edges, presenting relationship between two individuals\\
      \hline
    \end{tabular}%
  \label{tab:comm_notations}%
 \centering
\end{table}%

We assume that there is only one sample per node, i.e., $n=1$. However, the proposed models can be extended to the case of $n>1$. We further assume that the graph $(\mathcal{V}, \mathcal{E})$ is unweighted, noting that the models can be extended to weighted graphs. In the following subsections we occasionally drop indices $i$ and $t$ depending on the context to simplify notation. We denote with $\theta := \{W,b\}$ the set of model coefficients $W_{t}$, $b_{it}$, $\forall i,t$.

Let $h_{it}(x_i):= W_{t}^T x_i + b_{it}$, we consider the probability distribution for the choice of individual $i$ in class $t$, as below: 
\begin{equation}\label{eq:choice_model}
P(Y_{it}=y_{it}) = \frac{1}{1+e^{-y_{it}h_{it}(x_i)}}
\end{equation}
where $y_{it}$ can take values of $1$ or $-1$. Note that the following two state-of-the-art discrete choice models follow this probability distribution: (1) logit discrete choice model, which is proved to be equivalent to the logistics regression model \cite{DCA}, where $t\equiv 1$ and $b_{it} = b_{jt}, \forall \{i,j\} \in \mathcal{V}^2$; (2) latent class model, where $t > 1$ and $b_{it} = b_{jt}, \forall \{i,j\} \in \mathcal{V}^2,\forall t$. We also define our models with the same probability distribution according to Equation (\ref{eq:choice_model}): (1) LLGR model where $t\equiv 1$, and $b_{it}$ are not constant; (2) LCGR model where $t\equiv 1$, and $b_{it}$ are not constant.

\subsection{Local logistic graph regularization (LLGR)}\label{subsec:soc_log}
Choice model specified by Eq.~\eqref{eq:choice_model} includes several known models such as logistic regression, where there is no latent class, so $t$ is removed, $K=1$ and $y_i$ follows a Bernoulli distribution given $x_i$. To incorporate the social aspect in logistic regression one assumes that the parameters $b$ follow an exponential family parametrized with the given graph
\begin{align*}
P(b) \propto  \prod_{(i,j)\in \mathcal{E}} e^{-\lambda (b_{i}-b_{j})^2 },
\end{align*}
where $\lambda\in\mathbb{R}$ is a hyper-parameter. This model is usually trained by using a maximum log-likelihood estimator which reduces to the following regularized logistic regression problem
\begin{equation*}
\theta^* := \argmin_{\theta} \sum_{i=1}^N \log \left(1+e^{-y_{i}h_i(x_i)}\right)+ \lambda \sum_{(i,j)\in \mathcal{E}} (b_{i}-b_{j})^2,
\end{equation*}
where $h_i(x_i):= W^{T}x_i +b_i$. Notice that the social information, i.e., edges $\mathcal{E}$, appears as Laplacian regularization for the coefficients $b$. 

\subsection{Latent class graph regularization (LCGR)}\label{subsec:reg_latent}
The LCGR model is an extension to LLGR model with latent classes, $K>1$. In this model, $y_{it}$ follows a Bernoulli distribution given $x_i$ and $z_i=t$. To incorporate social information, we assume that latent class variables $z_i$ are distributed based on the following exponential family parametrized by the given social graph
\begin{equation}\label{eq:MRF}
P(z;b) \propto  \prod_{(i,j)\in \mathcal{E}} \exp\left(-\lambda \sum_{t=1}^{K} (b_{it}-b_{jt})^2  \mathbf{1}(z_{i} = z_{j} =t  )\right),
\end{equation}
where $b$ represents the collection of coefficients $b_{it}$ $\forall i,t,$ which are the parameters of the distribution.

Note that this specification allows utilizing social structures by introducing
 (1) continuous latent variables $b_{it}$ defined in graph regularization; (2) discrete latent variables $z_{i}$ defined in the above Markov Random Field. For continuous latent variables $b_{it}$ in this model, we assume that each individual has its own local coeffficient $b_{it}$ for each class. Notice that this model does not penalize different coefficients $b_{it}$ among connected individuals in different classes. This is because we assume that connected individuals in different classes should have independent linear classifiers. For discrete latent variables $z_{i}$ in the common latent class models, they are independent and identically distributed following the multinomial distribution, i.e., $z_{i} \sim \mbox{Mult}(\pi,1)$ $\forall i$, where $\pi$ is the probability of success. However, in our specification, hidden variables $z$ are correlated and are not necessarily identically distributed. Hence the continuous latent variables $b$ and the discrete latent class variable $z$ in our model can better model the observed choice processes, through which we can improve the model performance compared with the state-of-the-art models.

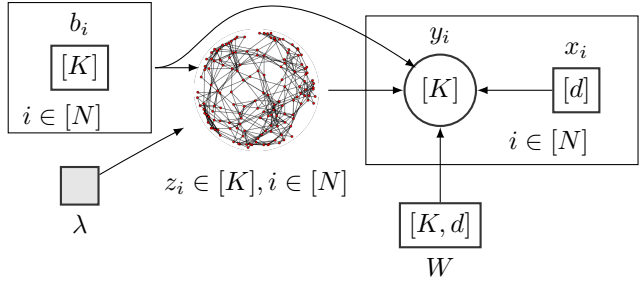
\begin{figure}
\centering
\hspace{-0.8cm}
\begin{tikzpicture}
\tikzstyle{main}=[circle, minimum size = 5mm, thick, draw =black!80, node distance = 7mm]
\tikzstyle{main2}=[circle, minimum size = 5mm, thick, draw =black!80, node distance = 10mm]
\tikzstyle{main3}=[minimum size = 5mm, thick, node distance = 10mm]
\tikzstyle{main4}=[minimum size = 5mm, thick, node distance = -6mm]

\tikzstyle{connect}=[-latex, thick]
\tikzstyle{box}=[rectangle, draw=black!100]
  \node[main, rectangle] (b) [label=above:$b_i$] {$[K]$};
  \node[main3] (z) [label=below:{$z_i\in[K], i\in[N]$}] [right= of b] [yshift=-0.3cm]{\includegraphics[width=.1\textwidth]{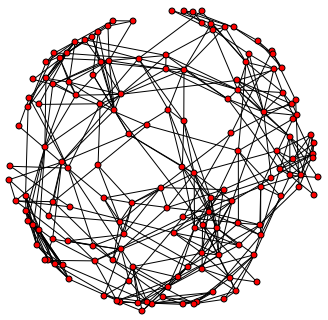}};
  \node[main2] (y) [label=above:$y_i$] [right= of z]{$[K]$};
  \node[main2, rectangle] (x) [label=above:$x_i$] [right= of y]{$[d]$};
  \node[rectangle, inner sep=0mm, fit= (y) (z),label=below right:{$i\in[N]$}, xshift=13mm, yshift=+5.5mm] {};
  \node[rectangle, inner sep=5.0mm,draw=black!100, fit= (y) (x)] {};
  \node[main2, rectangle] (w) [label=below:$W$] [below= of y]{$[K,d]$};
  \node[main, rectangle, fill = black!10, node distance = 10mm] (lambda) [label=below:$\lambda$] [below= of b]{};
  \node[rectangle, inner sep=5.6mm,draw=black!100, fit= (b)] {};
  \node[rectangle, inner sep=0mm, fit= (b),label=below right:{$i\in[N]$}, xshift=-12mm, yshift=-0.5mm] {};
  
  \draw[-latex] (1,0) -- (1.59,0);
  \draw[-latex] (1,0) .. controls (1.8,0) and (1.8,1.8) .. (4.47,0);
  \draw[-latex] (lambda) to[bend left=1] node[above] {} (z);
  \draw[-latex] (z) to[left=10] node[above] {} (y);
  \draw[-latex] (x) to[left=10] node[above] {} (y);
  \draw[-latex] (w) to[left=10] node[above] {} (y);
  
\end{tikzpicture}
\caption{Graphical model representation for social logistic regression models with latent variables. 
We use a modified plate notation to represent conditional dependence among random variables and dependence on parameters.
In particular, random variables are represented using circles and their number is shown in brackets inside the circle, i.e., $y_i$ corresponds to $K$ variables. Parameters are represented in rectangles and their sizes are shown in brackets with two components, i.e., $W$ corresponds to $K\times d$ coefficients. Data are shown in rectangles and their size in brackets, i.e., $x_i$ corresponds to $d$ features. There are $N$ nodes in the graph and each node corresponds to a random variable $z_i$ which takes values 
in $[K]:=\{1,\dots,K\}$. The hyper-parameter $\lambda$ is represented using a grey rectangle.}
\label{fig:graph_model}
\end{figure}

A graphical interpretation of this model is given in Figure \ref{fig:graph_model}. The resulting model can be trained using maximum likelihood and the Expectation-Maximization (EM) algorithm; details are discussed in Section \ref{sec:param_est}.

%% file: learning.tex
\section{Parameter Estimation}
In section, we focus on the parameter estimation algorithms for both the LLGR model and the LCGR model.
\label{sec:param_est}
\subsection{Local logistics graph regularization: ADMM}\label{subsubsec:scalmstep}
In this section we discuss how to minimize in distributed manner the negative expected log-likelihood function of the LLGR model. 
Following the work of \cite{hallac2015network} that applies ADMM to network lasso method, we extend it by allowing both local $b$ and global variables  $W$ on nodes. Let 
\begin{align}\label{eq:Qt}
Q(\theta;x,y)& := \sum_{i \in \mathcal{V}}\log \left(1+e^{-y_ih_{i}(x_i)}\right)  +\lambda \sum_{(i,j)\in \mathcal{E}} (b_{i}-b_{j})^2 
\end{align}
be the objective function, where $\theta$ represents the collection of parameters $W$ and $b$ and $h_{i}(x_i):= W^{T}x_i +b_{i}$. To minimize \eqref{eq:Qt} using ADMM, we introduce a copy of $b_{i}$ denoted by $z_{ij}$, $\forall i$, and a copy of $W$ denoted by $g_i$, $\forall i$, then we have:
\begin{equation}\label{eq:admm_prob}
\begin{array}{ll}
\underset{W,b}{\min} &  \sum\limits_{i \in \mathcal{V}} \log \left(1+e^{-y_i(g_i^Tx_i +b_{i})}\right) + \lambda \sum\limits_{(i,j)\in \mathcal{E}}  (z_{ij}-z_{ji})^2  \\
\text{s.t:}.        &  b_{i} = z_{ij} \ j\in \mathcal{N}(i), \ \forall i\\
				& W = g_i,  \ \forall i,
\end{array}
\end{equation}
where $\mathcal{N}(i)$ are the adjacent nodes of node $i$. By introducing copies for $b_{i}$, $\forall i$ we dismantle the sum over edges into separable functions. Additionally, 
by introducing copies for $W$, we dismantle the sum over the nodes for the logistic function. Then by relaxing the constraints we can make the problem \eqref{eq:admm_prob}
separable, which allows for distributed computation. 

We define the augmented Lagrangian below, where $u$ and $r$ are the dual variable and $\rho_1$ and $\rho_2$ are the penalty parameters.
\begin{align*}
&L_{\rho}(W,b,g,z,u,r) := \sum\limits_{i \in \mathcal{V}} \Big\{ \log \left(1+e^{-y_i(g_i^Tx_i +b_{i})}\right) \\ 
			    	 &+ \frac{\rho_1}{2} \Big(\left\Vert r_i\right\Vert^2_2   + \left\Vert W-g_i+r_i\right\Vert^2_2\Big) \Big\} +  \lambda \sum\limits_{(i,j)\in \mathcal{E}} \Big\{q_{ij} (z_{ij}-z_{ji})^2 \\
			     &+ \frac{\rho_2}{2}\Big(\left\Vert u_{ij}\right\Vert^2_2     + \left\Vert u_{ji}\right\Vert^2_2 + \left\Vert b_{i} - z_{ij}+u_{ij}\right\Vert^2_2 + \left\Vert  b_{j} - z_{ji}+u_{ji}\right\Vert^2_2 \Big)\Big\}.  
\end{align*}
The resulting ADMM algorithm is presented in Algorithm \ref{alg:admm}, where we set\\ $f(z_{ij}, z_{ji}):= L_{\rho}(W^{k+1},b^{k+1},g^{k+1},(z_{ij},z_{ji},z_{(ij)^c}^{k}),u^{k},r^{k})$.
Notice that the subproblems in Step $4$ do not have closed form solutions. However, they can be solved efficiently using an iterative algorithm since they are univariate problems that depend only on $x_i$ and not all data. Similarly, the subproblems in Step $5$ do not have closed form solution, but they have only $d$ unknown variables and depend only on $x_i$ and not all data. 
Moreover, Step $6$ has a closed form solution, which corresponds to solving a $2\times 2$ linear system. 
Observe that the ADMM algorithm \ref{alg:admm} can be run in a distributed setting by distributing the data among processors, because within each iteration, the computation of the value update of  each node and edge are independent.
\begin{algorithm}
\caption{ADMM for Problem \ref{eq:admm_prob}}
\label{alg:admm}
\begin{algorithmic}[1]
\Initialize{$k\gets 0$, $W^{k}$, $b^{k}$, $g^{k}$, $z^{k}$, $u^{k}$ and $r^{k}$} 
\Repeat
\State Set $W_t^{k+1} = \frac{1}{N}\sum_{i=1}^N (g_i^{k} - r_i^k)$
\State $b^{k+1}_{it} := \arg\min\limits_{b_{i}} \ L_{\rho}(W^{k+1}_t,b_{i},g^{k},z^{k},u^{k},r^{k})$ $\forall i \in \mathcal{V}$
\State $g^{k+1}_{i} := \arg\min\limits_{g_{i}} \ L_{\rho}(W^{k+1},b^{k+1},g_i,z^{k},u^{k},r^{k})$ $\forall i \in \mathcal{V}$
\State $z^{k+1}_{ij},z^{k+1}_{ji} =  \arg\min\limits_{z_{ij},z_{ji}} \ f(z_{ij}, z_{ji})$ $\forall (i,j) \in \mathcal{E}$
\State Set
\vspace{-0.2cm}
\begin{align}\nonumber
r^{k+1}_i    &=  r^{k}_i + (W^{k+1}-g^{k+1}_i) \ \forall i\in\mathcal{V}\\\nonumber
u^{k+1}_{ij} &= u^{k}_{ij} + (b^{k+1}_{i}-z^{k+1}_{ij})\ \forall (i,j)\in\mathcal{E}\\\nonumber
u^{k+1}_{ji} &= u^{k}_{ji} + (b^{k+1}_{j}-z^{k+1}_{ji})
\end{align}
\vspace{-0.4cm}
\State $k\gets k + 1$
\Until{termination criteria are satisfied.}
\end{algorithmic}
\end{algorithm}
\vspace{5mm}

\begin{figure}
\centering
\includegraphics[scale=0.4]{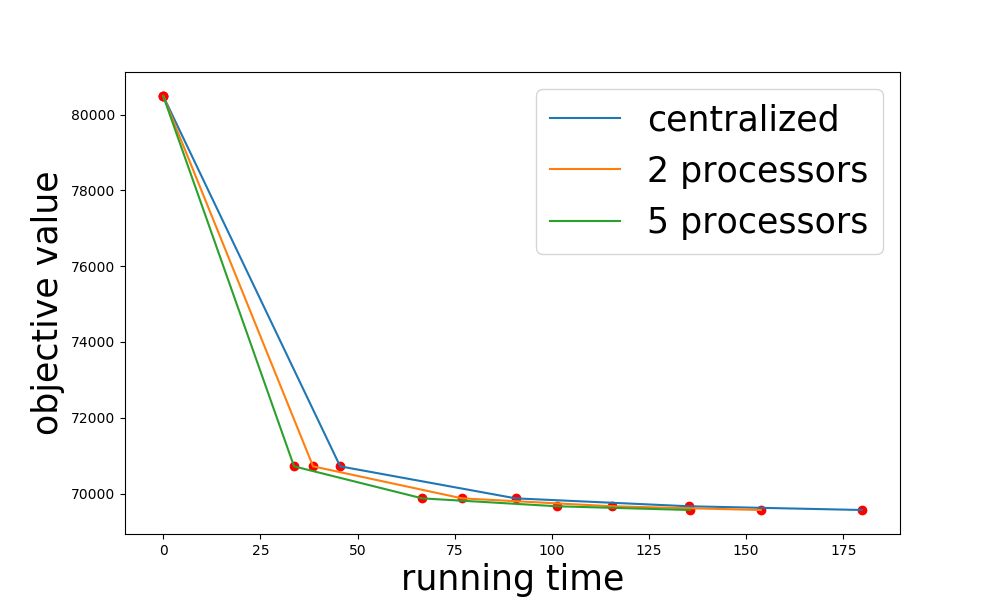}
 \caption{Illustration of the performance of ADMM algorithm with different number of processors. Each node represents the running time and objective value of a iteration in an experiment}
  \label{fig:M_timing}
\centering
\end{figure}
Figure \ref{fig:M_timing} demonstrates how distributing the data among processors can speed up the convergence of the ADMM algorithm. In this experiment, we randomly generate a binomial graph with 100k nodes, 500k edges. Then we randomly generate the feature matrix and response vector for the graph. We test the model on a server with 12 processors, and we use the multiprocessing package from Python to control the number of processors used in the parallel computing paradigm. As can be seen from Figure \ref{fig:M_timing}, distributing the computation among processors can greatly reduce the running time. Note that the running time does not decrease proportionally to the number of processors. It is because Python multiprocessing module is used here, but it takes time for process to communicate with the memory, and only the step for updating $b_{it}$ is parallelled.

\subsection{Latent Class Graph Regularization: Monte Carlo EM}\label{MCEM}
Generally, graphical models with categorical latent variables can be solved using the expectation maximization (EM) algorithms. However, correlations among latent variables imposed by the social graph do not allow exact calculation of posterior distributions in the E-step using standard EM approaches. Instead, an approximate calculation of the E-step using Monte Carlo EM (MCEM)~\cite{neath2013convergence,LC12,DLM99,WT90} is employed. It is a modification of the original EM algorithm where the E-step is conducted approximately using a Monte Carlo Markov Chain (MCMC) algorithm. The details for each step of the MCEM algorithm for the proposed models are provided in the following subsections.
\begin{algorithm}
\caption{MCEM algorithm for LCGR}
\label{alg:MCEM}
\begin{algorithmic}[1]
\Inputs{$\left(x_i,y_i\right)$, $i=1\dots N$}
\Initialize{$\theta^{0}:=\{W^{0},b^{0}\}\gets$ arbitrary value, $k\gets 0$} 
\Repeat
\State $\textbf{E-step:}$ (Subsection \ref{subsubsec:estep}) 
\State Calculate approximate node posterior \\for each node $i\in[N]$ 
	  \hspace{0.39cm} 
	  $$q(z_i=t):=P(z_{i}=t|y_{i},x_i;b^{k})$$,
	  \hspace{0.39cm} and for each edge $(i,j)\in\mathcal{E}$, the edge posterior 
	  $$q(z_i=z_j=t):=P(z_{i}=t,z_{j}=t|y_{i},y_{j},x_i,x_j;b^{k})$$\\
	  \hspace{0.39cm} by using the MCMC sampling.
\State $\textbf{M-step:}$ (Subsection \ref{subsubsec:mstep}) 
\State Solve the optimization problem
 	  $$\theta^{k+1} := \argmin_{\theta} Q(\theta;x,y),$$
	  \hspace{0.39cm} where $Q(\theta;x,y)$ is defined at \autoref{eq:Q}.
\State $k\gets k + 1$
\Until{termination criteria are satisfied.}
\end{algorithmic}
\end{algorithm}
\vspace{-5mm}

\subsubsection{Expectation step}\label{subsubsec:estep}
In this step the objective is to compute the marginal posterior distribution for nodes and edges, which will be used in the M-step to calculate 
the negative expected log-likelihood function. 
Refer to the Appendix for derivation of negative expected log-likelihood, which reveals the need for calculation of marginal posterior distributions.

In particular, for the E-step, one needs to calculate the following node marginal posterior probability
\begin{equation}\label{eq:node_prob}
P(z_{i}=t|y_{i},x_i;\theta) = \frac{P(y_{i}|x_i,z_{i}=t;\theta)  P(z_{i}=t;b)}{ \sum\limits_{s=1}^{K} P(y_{i}|x_i,z_{i}=s;\theta) P(z_{i}=s;b) }
\end{equation}
and the following edge posterior probability
\begin{align}\label{eq:edge_prob}
&P(z_{i}=t,z_{j}=t|y_{i},y_{j},x_i,x_j;\theta)\\\notag
&=\frac{P(y_{i},y_{j}|x_i,x_j,z_{i}=t,z_{j}=t;\theta)  
P(z_{i}=t,z_{j}=t;b)}{ \sum\limits_{m,q=1}^{K}  P(y_{i},y_{j}|x_i,x_j,z_{i}=m,z_{j}=q;\theta) P(z_{i}=m,z_{j}=q;b) },
\end{align}
where $\theta$ represents the collection of parameters $W$ and $b$.
For small graphs, we can approximate the above distributions using standard MCMC algorithms. For large graphs, please refer to the Scalability of the E-step Subsection \ref{subsubsec:scalestep} in the Appendix.

\subsubsection{Maximization step}\label{subsubsec:mstep}
Let us denote with $q(z_i=t) = P(z_{i}:=t|y_{i},x_i;\theta)$ and $q(z_i=z_j=t):= P(z_{i}=t,z_{j}=t|y_{i},y_{j},x_i,x_j;\theta)$ the marginal posterior distributions.
The M-step of the EM algorithm requires minimizing the negative expected log-likelihood function
\begin{align}\label{eq:Q}
Q(\theta;x,y) :=& \sum_{i \in \mathcal{V}} \sum_{t=1}^K q(z_i=t) \log \left(1+e^{-y_i h_{it}(x_i)}\right) \\ 
\notag
 & + \lambda\sum_{(i,j)\in \mathcal{E}} \sum_{t=1}^{K} (b_{it}-b_{jt})^2 q(z_i=z_j=t),
\end{align}
where $\theta$ represents the collection of parameters $W$ and $b$ and $h_{it}(x_i):= W_t^{T}x_i +b_{it}$.
Derivation of this function is given in Subsection \ref{subsec:elog} in the Appendix. For small graphs, standard convex optimization solvers can be used. For large graphs, please refer to Section \ref{subsubsec:scalmstep} where we discuss how we can maximize the expected log-likelihood efficiently with a distributed algorithm for LLGR. The objective function of the E step of LCGR model is the weighted version of the objective function of the LLGR model.

We now comment briefly on the theoretical asymptotic convergence of MCEM to a stationary point of the likelihood function.
Convergence theory of MCEM in \cite{neath2013convergence,DLM99} states that if standard MCMC is used in E step and that the MCMC sample size increases deterministically across MCEM iterations, then MCEM converges almost surely. We also consider using parallell block MCMC for large graphs (please refer to the Scalability of the E-step Subsection \ref{subsubsec:scalestep} in the Appendix), then a consequence of blocking of latent variables for the MCMC algorithm is that asymptotic convergence of MCEM is not guaranteed anymore. However, in practice, MCEM is often terminated without even knowing if the algorithm converges to an accurate solution. See for example Section $5$ in \cite{neath2013convergence} and references therein about arbitrary termination criteria of MCEM. Therefore, we consider that the parallelism of block MCMC Algorithm in Subsection \ref{subsubsec:scalestep} of the Appendix, offers a trade-off between convergence and computational complexity per iteration by controlling the number of blocks, which in practice can speed up each iteration of the MCEM algorithm significantly.

%% file: Experiments.tex
\section{Experiments}\label{sec:experiments}
In this section, we analyze the empirical performance of the proposed social models on a range of datasets. We outline practical recommendations and illustrate examples where the proposed model is most suitable. The number of iterations of Gibbs sampler in the E-step grows with the number of iterations of the MCEM algorithm. The M-step is implemented using ECOS solver~\cite{domahidi2013ecos} embedded in CVXPY~\cite{diamond2016cvxpy} for $W$ updates, bisection line search for $b$ updates, within ADMM iterations.

\input{Experiments_1}
\input{Experiments_2}
\input{Experiments_3}

%% file: Experiments_1.tex
\subsection{Illustrative Synthetic Data}
\label{sec:experiment1}
We demonstrate that when graph structure and label assignment satisfy certain conditions, our LLGR and LCGR models performs better than other models without social information. We empirically illustrate the impact of graph structure by varying the connectivity between different classes in the graph, and illustrate the impact of label assignment by varying the discrepancies of labels in communities of the graph.
\subsubsection{Varying connectivity between classes}
\begin{figure}
\centering
\begin{subfigure}{.25\textwidth}
  \centering
  \includegraphics[scale=0.20]{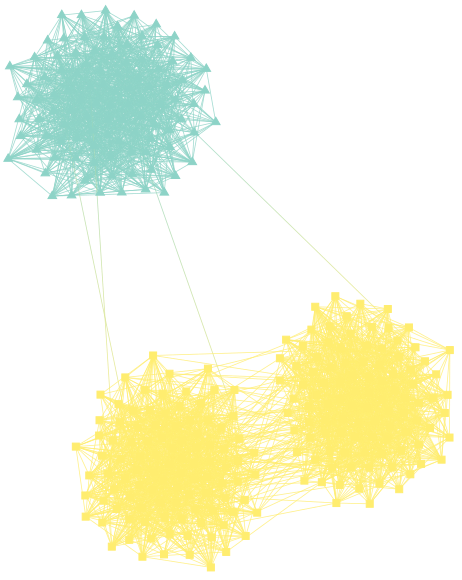}
  \caption{$\beta=10^{-4}$}
  \label{fig:synthetic_graph_3_1}
\end{subfigure}%
\begin{subfigure}{.25\textwidth}
  \centering
  \includegraphics[scale=0.20]{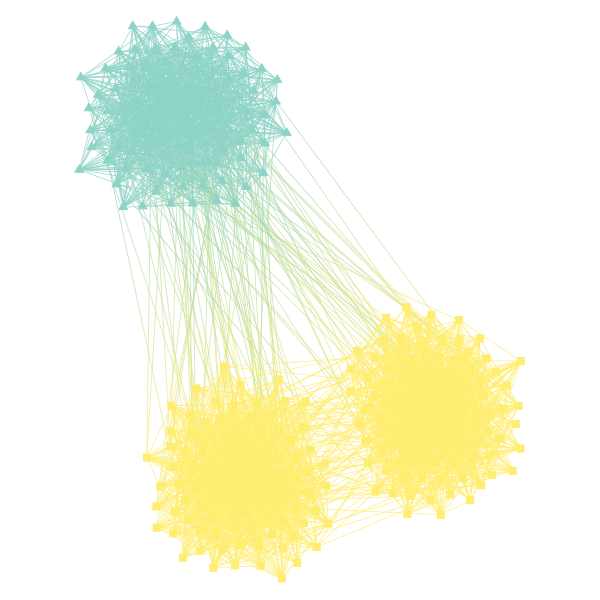}
  \caption{$\beta=10^{-2}$}
\end{subfigure}
\begin{subfigure}{.25\textwidth}
  \centering
  \includegraphics[scale=0.20]{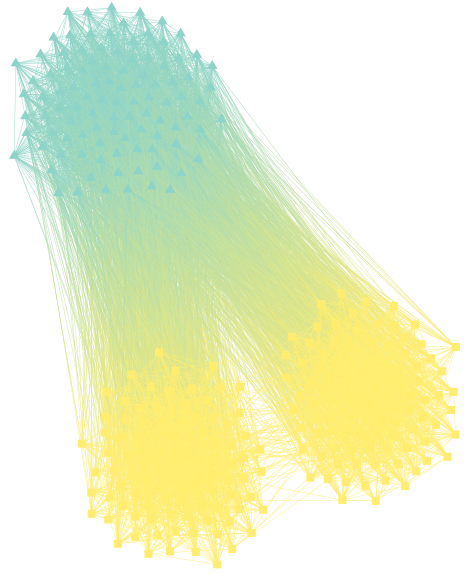}
  \caption{$\beta=10^{-1}$}
 \label{fig:synthetic_graph_3_3}
\end{subfigure}
\caption{The nodes with square shape and yellow color correspond to class $t=1$. The nodes with triangle shape and turquoise color correspond to class $t=2$. The larger $\beta$ the more edges among nodes with different class.}
\label{fig:synthetic_graph_3}
\end{figure}

We consider $N=300$ individuals, and use two different Gaussian distribution to  generate feature vector for each individual. Then we randomly split the individuals into three communities with the same size. We assume that there are two classes, shown in blue and yellow in \autoref{fig:synthetic_graph_3}. Notice that the feature vectors are assigned to communities regardless of their Gaussian distribution and labels are set based on the classes, which correspond to communities. Therefore, the labels are in align with the graph structure but not in align with the feature space. We set the probability of two individuals that are in the same community to get connected to $0.2$, and the probability of two individuals that are in the same class but not in the same community to get connected to $0.01$. Then we vary parameter $\beta$, the probability of two individuals in different classes to get connected.

\autoref{fig:synthetic_graph_3} shows the graph structure when $\beta = 10^{-4}$, $\beta = 10^{-2}$ and $\beta=10^{-1}$. Notice that the larger $\beta$ is, the more edges among the communities belong in different classes. 
Table \ref{tab:synthetic_beta} shows the prediction result of four models as a function $\beta$. Notice that since feature vectors and labels are not changed as $\beta$ changes, the prediction of logistic regression and logistic regression with latent class remains constant at $62\%$. The reason that these models perform poorly is because the labels are not separable given the feature vectors $x_i$ only. Observe in Table \ref{tab:synthetic_beta} that when $\beta$ is as small as $10^{-4}$, which means that individuals in different classes are very unlikely to get connected, 
see Figure \ref{fig:synthetic_graph_3_1}, the prediction result of the proposed social models is larger than $80\%$. On the other hand, when $\beta$ becomes larger, the prediction of the social models is declining. However, as long as $\beta < 0.1$, the proposed LCGR model performs better than logistic regression and logistic regression with latent class models. When $\beta=0.1$ the social models has the same prediction performance as the logistic regression and latent class models. This is because the classes are not clearly separable on the graph, see Figure \ref{fig:synthetic_graph_3_3}.

\vspace{-3 mm}
\begin{table}[htbp]
  \centering
  \caption{Prediction results on a randomly chosen test set of $50$ individuals when $\beta$ is varied, i.e., connectivity between classes. For all models the regularization parameter $\lambda$ which corresponds to the best prediction result out of a range of parameters is chosen.}
    \begin{tabular}{cccccc}
    \toprule
    & \multicolumn{5}{c}{$\beta$}\\
    model & $10^{-4}$&$10^{-3}$ &$5\times 10^{-3}$ & $10^{-2}$ & $10^{-1}$\\
    \midrule
    logistic reg. & 62\% &62\% & 62\%&62\% &62\% \\
    log. reg. lat. class &62\% &62\% &62\%& 62\%& 62\%\\
    LLGR & 80\% & 62\% & 62\%& 62\%& 62\%\\
    LCGR & 88\% & 82\% & 64\%& 62\%& 62\%\\
    \bottomrule
    \end{tabular}%
  \label{tab:synthetic_beta}%
\end{table}

\subsubsection{Varying choice preference parameters}
An ideal scenario for the proposed social models is when classes correspond to communities of the given graph and when the labels $y_i$ are clustered according to the classes. However, labels $y_i$ might be misplaced in wrong classes. We study how the preference difference between classes affects the performance of the proposed model.
 
For this experiment individual feature vectors are generated from a Gaussian distribution with sample size $N=200$. We randomly split this set of individuals into two parts with same size, and each part represents a class where individuals share the same parameters $W$. Assume that $W_i$ is the weight corresponding to the $i$th group, and $W_1=-W_2$. 
For each individual $j$, $b_j$ is sampled from the same Gaussian distribution. For the graph setting, we set the probability of people in the same class to be connected as $0.2$, and the probability of people in different classes to be connected as $10^{-4}$. This way, we ensure that classes correspond to communities.
 
\begin{figure}
\centering
\begin{subfigure}{.25\textwidth}
  \centering
  {\includegraphics[scale=0.20]{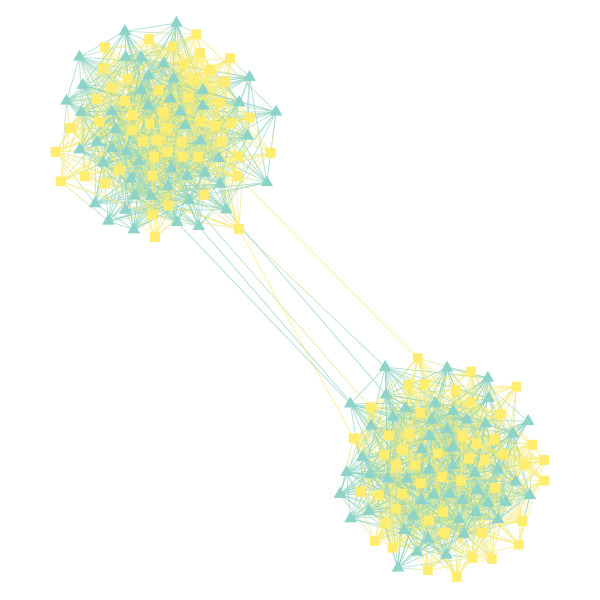}}
  \caption{$\|W_1\|_2=0$}
\end{subfigure}%
\begin{subfigure}{.25\textwidth}
  \centering
  {\includegraphics[scale=0.20]{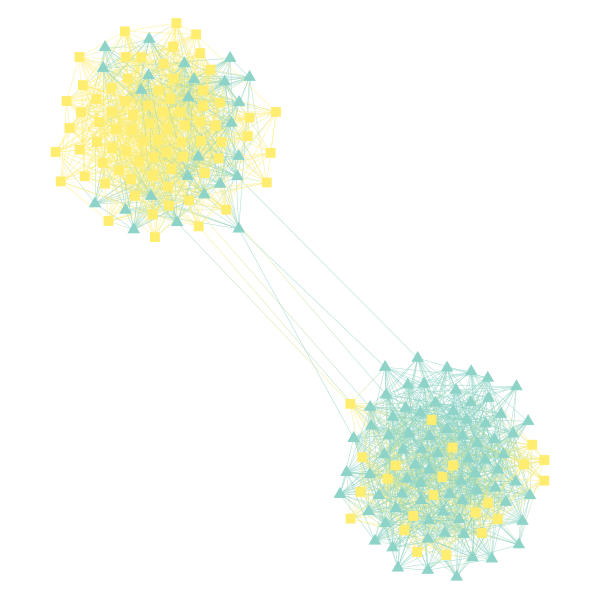}}
  \caption{$\|W_1\|_2=4$}
\end{subfigure}
\begin{subfigure}{.25\textwidth}
  \centering
  {\includegraphics[scale=0.20]{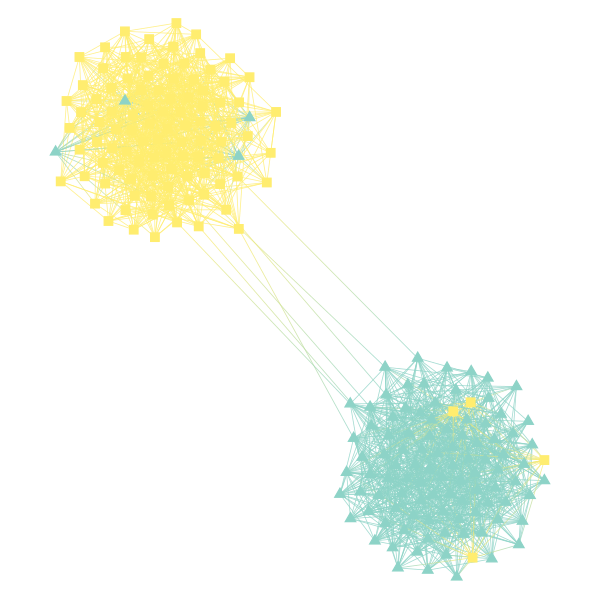}}
  \caption{$\|W_1\|_2=6$}
\end{subfigure}
\caption{Three synthetic examples showing the influence of parameter $W_1$ on class preference. The nodes with square shape and yellow color correspond to choice $y_i=1$. The nodes with triangle shape and turquoise color correspond to choice $y_i=-1$.}
\label{fig:synthetic_graph_w}
\end{figure}

Based on the data generation process, by tuning $\|W_1\|$, we are able to get full control of preference difference among individuals in the two classes. When $\|W_1\|$ becomes larger, preference difference becomes larger as well. As we see in \autoref{fig:synthetic_graph_w}, when $\|W_1\|$ becomes larger, more individuals in class one have $y_i =1$ (i.e., yellow squares) and more individuals in class two have have $y_i =1$ (i.e., turquoise triangles). When $W_1=0$, around half of the individuals in both classes have $y_i =1$, the other half $y_i = -1$, which means that there is no preference difference between the two classes.
Prediction results for this experiment are shown in Table \ref{tab:synthetic_w}.
\vspace{-3 mm}
\begin{table}[htbp]
  \centering
  \caption{Prediction results on a randomly chosen test set of $50$ individuals when $\|W_1\|_2$ is varied. For all models the regularization parameter $\lambda$ which corresponds to the best prediction result out of a range of parameters is chosen.}
    \begin{tabular}{ccccccc}
    \toprule
          & \multicolumn{6}{c}{$\|W_1\|_2$}\\
    model & $10$ & $5$ &$3$ & $2$ & $1$  & $0$\\
    \midrule
    logistic reg. & 48\% &44\% & 42\%&52\% &58\%  &52\%\\
    log. reg. lat. class &48\% &48\% &54\%&54\%
    &42\% &36\%\\
    LLGR and LCGR & 100\% & 100\% & 94\%& 86\%& 68\%  &36\%\\
    \bottomrule
    \end{tabular}%
  \label{tab:synthetic_w}%
\end{table}%
\vspace{-5 mm}

%% file: Experiments_2.tex
\subsection{Adolescent smoking}
This example uses a dataset collected by \cite{bush1997role}. This research program, known as the teenage friends and lifestyle study, has conducted a longitudinal survey of friendships and the emergence of the smoking habit (among other deviant behaviors) in teenage students across multiple schools in Glasgow, Scotland.

\subsubsection{Dataset}
\begin{figure}
\centering
\begin{subfigure}{.4\textwidth}
  \centering
  {\includegraphics[scale=0.38,trim=3.25cm 3cm 3cm 3cm,clip]{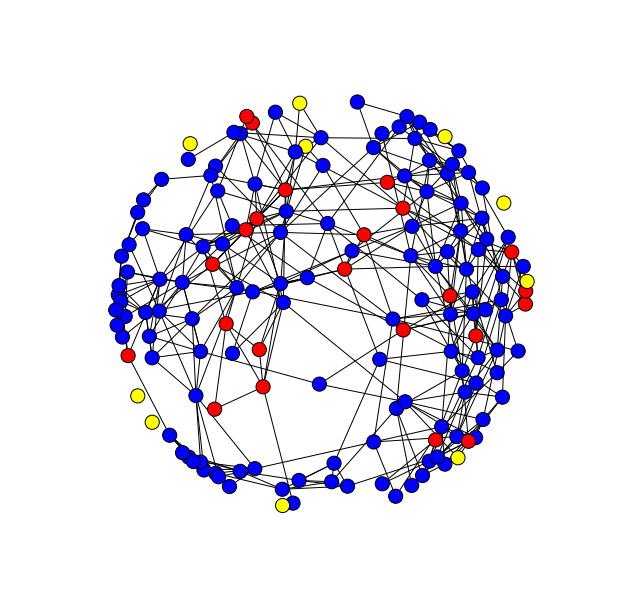}}
  \caption{Start of the study (Feb 1995):\\ 127 non-smokers (blue),\\ 23 smokers (red),\\ 10 unobserved (yellow)\\ 422 edges.}
\end{subfigure}%
\begin{subfigure}{.4\textwidth}
  \centering
  {\includegraphics[scale=0.4,trim=1cm 1cm 1cm 1cm,clip]{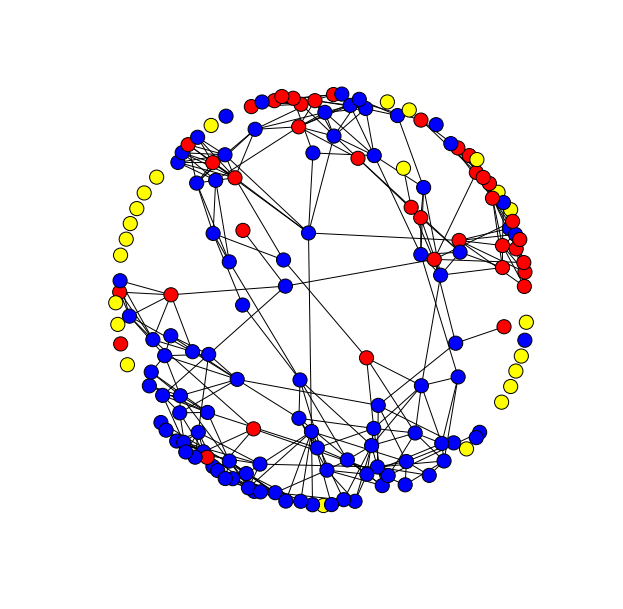}}
  \caption{End of the study (Jan 1997):\\98 non smokers (blue),\\ 39 smokers (red),\\23 unobserved (yellow) \\ 339 edges.}
  \label{fig:2years_2}
\end{subfigure}
\caption{Social graphs of student friendships and smoking behaviors within the 2 years period of the study.}
\label{fig:2years}
\end{figure}

Social graphs of $160$ students (shown in Figure~\ref{fig:2years}) within the same age range of 13-15 years is constructed following a surveyed evidence of reciprocal friendship, with an edge placed among individuals $i$ and $j$ if individual $i$ and individual $j$ named each other friends. We included five variables into the feature vector $x_i$: age; gender; money: indicating how much pocket money the student had per month, in British pounds; romantic: indicating whether the student is in a romantic relationship; family smoking: indicating whether there were family members smoking at home.
Notice that the feature vectors $x_i$ and the edges of the graph
are different at different timestamps.
The response variable $y$ represents the stated choice that whether a student smokes tobacco ($y_i=1$), otherwise $y_i=-1$. Note there are nodes with missing labels, but the graph structure should be intact for the parameter estimation of $b$. Therefore, we set $y_i = 0$ for these nodes, so that the corresponding $x_i$ are not used in the parameter estimation while keeping the graph structure unchanged.

\subsubsection{Models comparison}
We measured predictions on this dataset and report here 3-fold cross validation results of four models: i) logistic regression; ii) latent class logistic regression; iii) social logistics regression, see Subsection \ref{subsec:soc_log}; iv) social latent class logistics regression, see Subsection \ref{subsec:reg_latent}. Cross-validation process treats the removed nodes as nodes with missing labels, as described above. We consider $K=2$ latent classes in this experiment. The prediction performance is shown in Tables~\ref{tab:smoking9597}.

\begin{table}[htbp]
  \centering
  \caption{Adolescent smoking prediction accuracy, February 1995 and January 1997}
  \vspace{-0.3cm}
    \begin{tabular}{ccc}
    \toprule
    model & 1995 & 1997\\
    \midrule
    logistic regression & 81.1\% & 68.5\%\\
    latent class & 78.9\% & 72.1\%\\
    LLGR & 80.0\% & 76.9\%\\ LCGR & 82.2\% & 80.8\% \\    
    \bottomrule
    \end{tabular}%
  \label{tab:smoking9597}%
\end{table}%

High performance of logistic regression and latent class model for the beginning of the study (Table~\ref{tab:smoking9597}) indicates that the smoking preferences are largely defined by individual feature vector. Moreover, parameters in the underlying classes of the latent class model don't differ much. Social latent class model performs equally well. 
This difference grows significantly when a confounding variable of smoking in the family is removed from the feature list. 

Furthermore, by the end of the study (January 1997, Table~\ref{tab:smoking9597} column 2) one can see a significantly higher predictive accuracy of the LLGR and LCGR models. It may indicate that smoking within a certain group of teenagers have become a social norm (indicating the difference in offset parameters $b_{it}$), or that the response $y_i$ to independent factors $x_i$ within that group differs from the others as reflected by differences in $W_t$. 

Notice the significant decrease of prediction accuracy for non social models between column 1 and column 2 in Table \ref{tab:smoking9597}.
This is because at the beginning of the study (February $1995$)
less than $15\%$ of teenagers smoked, while at the end of the study (January $1997$) about $25\%$ of teenagers smoked.
This difference is likely due to social norms developed among teenagers that are captured by the graph and therefore missed by non social models.

\subsubsection{Parameter visualization}
We are going to illustrate how the specification of the proposed model allows in-depth exploration of the parameters to assist in making this type of conclusions. To that end, we are going to explore parameters $b_{it}$ and class membership probabilities across the regularization path of hyper-parameter $\lambda$.

\begin{figure}
\centering
\begin{subfigure}{.3\textwidth}
  \centering
  {\includegraphics[scale=0.3,trim=1cm 1cm 1cm 1cm,clip]{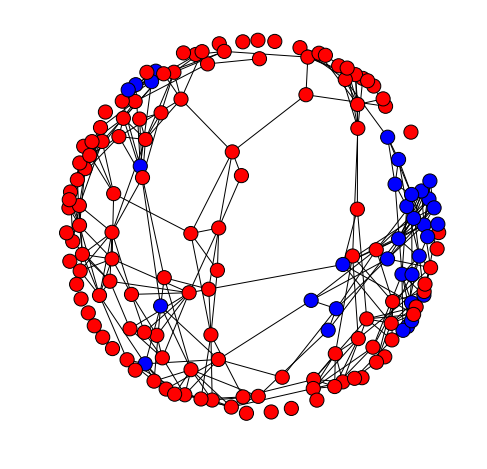}}
  \caption{$\lambda = 10$}
\end{subfigure}%
\begin{subfigure}{.3\textwidth}
  \centering
  {\includegraphics[scale=0.3,trim=1cm 1cm 1cm 1cm,clip]{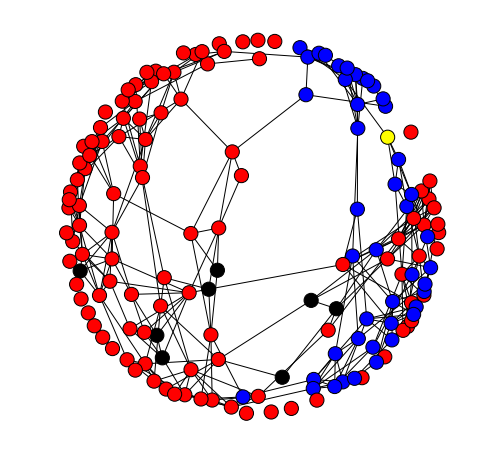}}
  \caption{$\lambda = 1$}
  \label{fig:class_vi2}
\end{subfigure}
\begin{subfigure}{.3\textwidth}
  \centering
  {\includegraphics[scale=0.3,trim=1cm 1cm 1cm 1cm,clip]{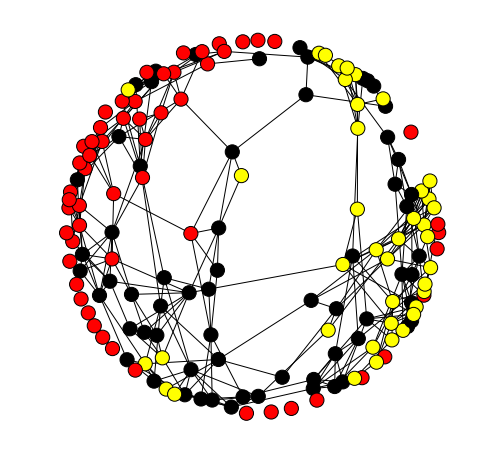}}
  \caption{$\lambda = 0.01$}
\end{subfigure}
\caption{Class membership probabilities estimated for the nodes for multiple values of $\lambda$. Blue, yellow, green, black and red colours correspond to probabilities about $0$, $(0,0.5)$, $0.5$, $(0.5,1)$ and $1$, respectively.}
\label{fig:class_vi}
 \end{figure}

\begin{figure}
\centering
\begin{subfigure}{.3\textwidth}
  \centering
  {\includegraphics[scale=0.3,trim=1cm 1cm 1cm 1cm,clip]{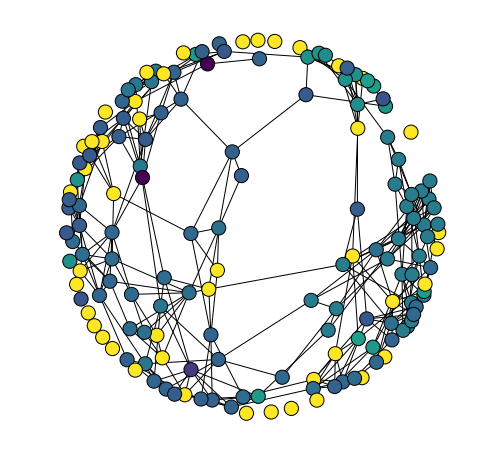}}
  \caption{$\lambda =10$}
\end{subfigure}%
\begin{subfigure}{.3\textwidth}
  \centering
  {\includegraphics[scale=0.3,trim=1cm 1cm 1cm 1cm,clip]{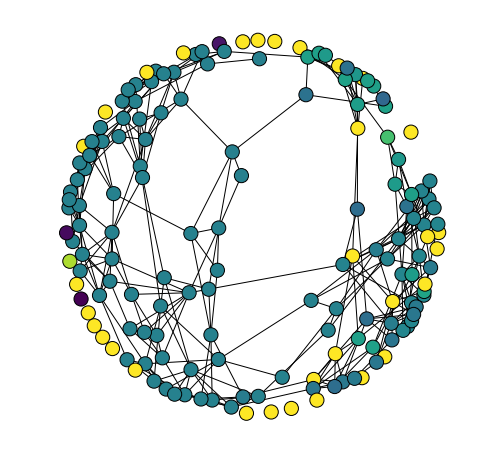}}
  \caption{$\lambda = 1$}
  \label{fig:b_vi2}
\end{subfigure}
\begin{subfigure}{.3\textwidth}
  \centering
  {\includegraphics[scale=0.3,trim=1cm 1cm 1cm 1cm,clip]{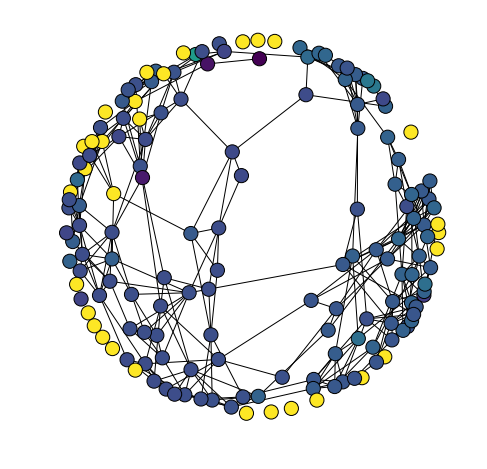}}
  \caption{$\lambda = 0.01$}
\end{subfigure}
 \caption{The values of $b_{it}$ estimated for the nodes for multiple values of $\lambda$. Lighter color corresponds to higher values.}
 \label{fig:b_vi}
 \end{figure}

The estimated class membership (the probability of being in a given latent class) on the graph is shown in Figure~\ref{fig:class_vi}. A visualization of estimated $b_{it}$ of one latent class for several values of $\lambda$ is shown in Figure~\ref{fig:b_vi}. 
When $\lambda=0.1$, smoking pattern begins to show across the graph, i.e., compare Figures \ref{fig:2years_2} and \ref{fig:b_vi2}.
Let us note that $\lambda=0.1$ value corresponds to the best prediction accuracy in our experiments for the proposed social latent class logistic regression model
for the smoke data at the end of the study (January $1997$). This 
is because the social latent class logistic regression model is able
to clearly distinguish a group of socially connected individuals within which the choice preferences towards smoking are higher.

When $\lambda=0.01$, $b_{it}$ are similar across the nodes. On the other hand, when $\lambda=10$, $b_{it}$ are not similar across the nodes. Although this is counter-intuitive since the graph regularization favors similar $b_{it}$ across nodes
for large values of $\lambda$, it is explained by the node and edge posterior distribution in the M-step which also controls regularization across nodes during the execution of the algorithm. 

%% file: Experiments_3.tex
\subsection{Online Retail Account Relation Graph Data}
In this section, we experimented our model on a real world online retail account relation graph from a leading commercial company in the US, and showed that our model excelled in fraud account prediction task when compared with other models. \textbf{More details on the data and the empirical results have been redacted until their release has been approved}

%% file: conclusion.tex
\section{Conclusions and Future Work}\label{sec:conclusions}
In this paper we introduced social graph regularization ideas into discrete choice models for user choice modeling. We proposed local logistics graph regularization (LLGR) method and latent class graph regularization (LCGR) model. We developed scalable parameter estimation method for LLGR model on large graphs  benefiting from recent advances in distributed optimization based on ADMM methods. Also we have developed, implemented, and explored parameter estimation algorithms that allow parallel processing implementation for both E- and M-steps of the Monte Carlo Expectation Maximization (MCEM) algorithms for LCGR model. In experimental evaluation, we have focused on investigating the usefulness of the models in revealing and supporting hypothesis in studies where not only predictive performance (that was found to be highly competitive), but also understanding social influence, is crucial. Our models can be directly applied to study social influence on revealed choices in large social graphs with rich node attributes. One challenge with extending our results is that such data are very rarely available in open access due to privacy issues.

%% file: appendix.tex
\appendix
\section{Improving E step}
\label{subsubsec:scalestep}
Here, we describe how to scale up the E step in Section \ref{subsubsec:estep}, which is important for MCEM method, as described in Section \ref{MCEM}.
In the LCGR model, we use a markov random field to model the joint distribution of latent class conditioned on local coefficients $b$. And we need to calculate the posterior node probabilities (Equation \eqref{eq:node_prob}) and edge probabilities (Equation \eqref{eq:edge_prob}) in the E step based on Equation ~(\ref{eq:MRF}). 
 Let us consider the case of two classes as an example. Assume the labels of the two classes are $1$ and $-1$, then we can simplify Equation~(\ref{eq:MRF}) to
\begin{align*}
P(z;b) \propto & \prod_{(i,j)\in \mathcal{E}} \exp((\theta_{ij}-\beta_{ij}) z_i/4 + (\theta_{ij}+\beta_{ij})z_j/4\\
&+ (\theta_{ij}
+\beta_{ij})z_i\times z_j/4 +  (\theta_{ij}+\beta_{ij})/4)
\end{align*}
where we have $\theta{ij}=(b_{ik}-b_{jk})^2, k=1$ and $\beta{ij}=(b_{ik}-b_{jk})^2, k=-1$, and $\exp((\theta_{ij}-\beta_{ij}) z_i/4 + (\theta_{ij}+\beta_{ij})z_j/4
+ (\theta_{ij}
+\beta_{ij})z_i\times z_j/4 +  (\theta_{ij}+\beta_{ij})/4)$ is the edge potentials. Thus, our model belongs to the standard pairwise Markov Random Field.

 Both variational methods and sampling-based methods are suitable for our problem setting. (1) For variational methods, exact inference of Markov Random Field using Bethe approximation \cite{willsky2008loop,wainwright2008graphical} can calculate distributions of nodes and edges. The approximate inference algorithms, e.g., mean field inference \cite{wainwright2008graphical} and loopy belief propagation (BP) \cite{murphy1999loopy}, although  exhibiting excellent empirical performance, $P(z_{i}=t,z_{j}=t;b)$ is not calculated. (2) For sampling-based approach, MCMC algorithms can calculate both the node and edge distributions.
 
 Calculating the marginal posterior probabilities Equation \eqref{eq:node_prob} and Equation \eqref{eq:edge_prob} is computationally expensive due to the marginal probabilities $P(z_{i}=t;b)$ and $P(z_{i}=t,z_{j}=t;b)$. This is because to calculate the latter two, we have to marginalize $N-1$ and $N-2$ latent variables, respectively. We chose sampling-based method and accelerated the computation by using a block MCMC sampling technique to compute $P(z_{i}=t;b)$ and $P(z_{i}=t,z_{j}=t;b)$. The algorithm uses a preprocessing step to partition the graph into $c$ disjoint communities. Then it runs an MCMC algorithm on each community/block in \textit{parallel} by ignoring the edges among the blocks.

\begin{figure}
\centering
\begin{subfigure}{.5\textwidth}
  \centering
  {\includegraphics[scale=0.38]{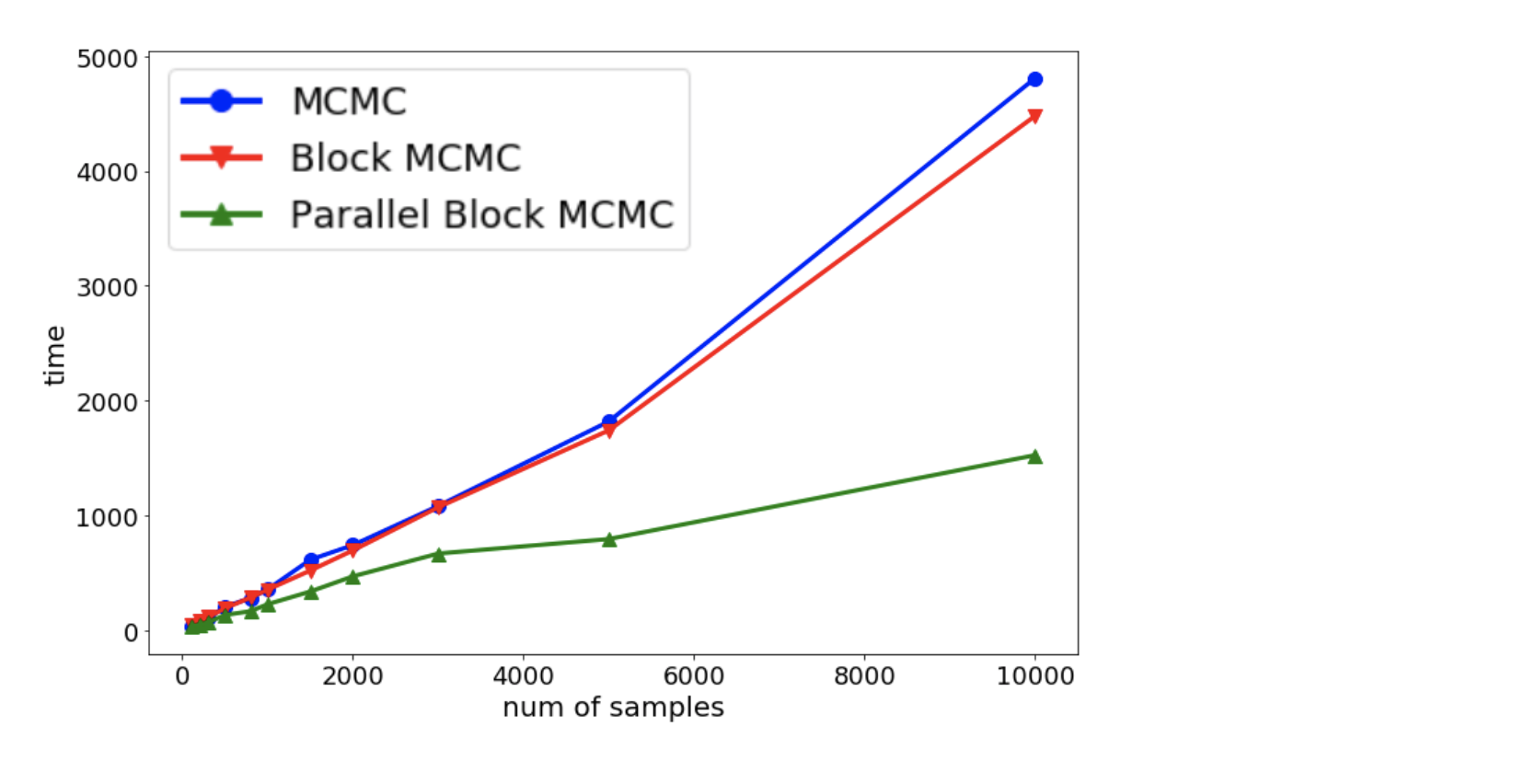}}
\caption{Parallel block MCMC computation time }
 \label{fig:E_timing_1}
\end{subfigure}
\begin{subfigure}{.4\textwidth}
  \centering
  {\includegraphics[scale=0.36]{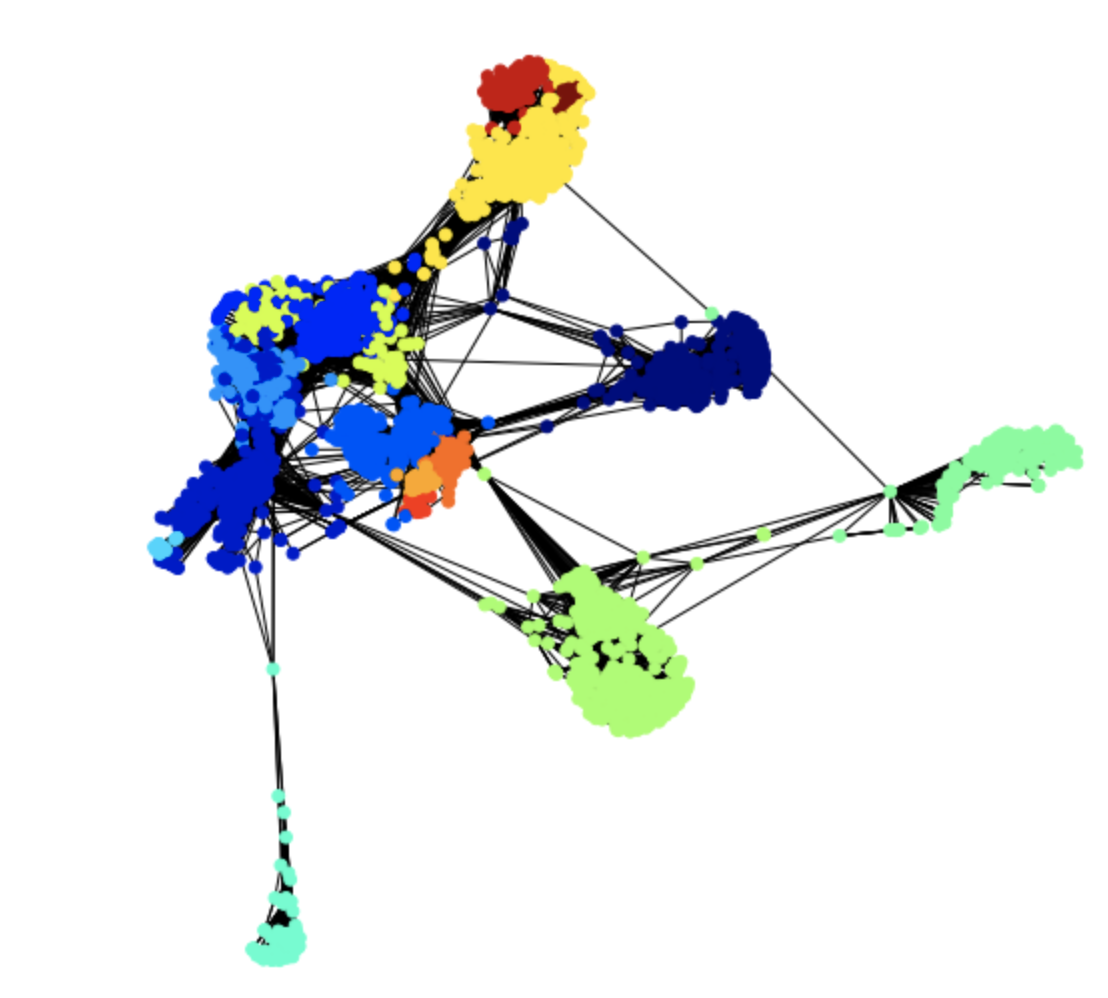}}
  \caption{Facebook Ego Network \\Community Detection Result}
  \label{fig:FB}
\end{subfigure}%
\caption{Left: Visualization of Facebook Ego Graph, where color represents community membership. Right: plot of number of samples VS running time, for standard MCMC, block MCMC and parallel block MCMC}
\end{figure}

Figure \ref{fig:E_timing_1} demonstrates how parallel block MCMC speed up the computation. We use the Facebook Ego Network Data as shown in Figure \ref{fig:FB}, where there are 4039 nodes and 88234 edges. We randomly generate the local b value of each class for all the nodes. We first run Louvian community detection algorithm and extract 10 communities from the graph. We then run MCMC/ Block MCMC and Parallel Block MCMC and time the code for each iterations. As can be seen from Figure \ref{fig:E_timing_1}, the performance of MCMC and Block MCMC are almost the same, because the only difference is that Block MCMC omits the edges between communities for the Gibbs update. Since the MCMC update of each community is independent from each other, we can adopt the parallel computing paradigm for Block MCMC. As we can see from Figure \ref{fig:E_timing_1}, Parallel Block MCMC performs much better than the other two. 
\section{Negative expected log-likelihood in Eq. (8)}\label{subsec:elog}

We denote with $q(z):= P(z | y,x;\theta)$ the posterior distribution, with $\sum_{z}$ the sum over all latent variables $z$, 
and $\theta$ represents the collection of parameters $W$ and $b$. Let $\mathbf{1}(z_{i}=t)$ be the indicator function, which is equal to $1$ if $z_i=t$.
We assume that $z$ follows
\begin{equation}
P(z;b) \propto  \prod_{(i,j)\in \mathcal{E}} \exp\left(-\lambda \sum_{t=1}^{K} (b_{it}-b_{jt})^2  \mathbf{1}(z_{i} = z_{j} =t  )\right),
\end{equation}
and 
\begin{align}\label{eq:p_y_x}
P(y_{i}|x_{i},z_{i}=t,\theta) = 1/(1+e^{-y_{i}h_i(x_i)}),
\end{align}
where $h_{it}(x_i):= W_t^{T}x_i +b_{it}$.
The derivation of the expected log-likelihood is shown below.
\begin{align*}
&\tilde{Q}(\theta;x,y) := \sum_z q(z) \log P(y, z | x; \theta)\\
&= \sum_z q(z) \log(P(y|x,z;\theta) P(z;b)) \\ 
&= \sum_z q(z) \log P(y|x,z;\theta) + \sum_z q(z) \log P(z;b)\\
&= \sum_z q(z) \log \left(\prod_{i \in \mathcal{V}}\sum_{t=1}^K P(y_{i}|x_{i},z_{i}=t,\theta)\mathbf{1}(z_{i}=t)\right) \\ 
&= \sum_z q(z) \sum_{i \in \mathcal{V}} \log \left(\sum_{t=1}^K P(y_{i}|x_{i},z_{i}=t,\theta)\mathbf{1}(z_{i}=t)\right) \\ 
&\hspace{5mm}- \lambda \sum_z q(z) \sum_{(i,j)\in \mathcal{E}} \sum_{t=1}^{K}  (b_{it}-b_{jt})^2  \mathbf{1}(z_{i} = z_{j} =t  ).
\end{align*}
We can exchange the sequence of $\log$ and $\sum_{t=1}^K$ because each node can only be in one class, thus we have
\begin{align*}
\tilde{Q}(\theta;x,y)
 & = \sum_z q(z) \sum_{i \in \mathcal{V}} \sum_{t=1}^K \mathbf{1}(z_{i}=t)  \log P(y_{i}|x_{i},z_{i}=t;\theta)
 - \sum_z q(z) \sum_{(i,j)\in \mathcal{E}} \sum_{t=1}^{K}  \lambda (b_{it}-b_{jt})^2  \mathbf{1}(z_{i} = z_{j} =t).
\end{align*}
Let's write the summation over $z$ inside the summation over vertices and the summation over latent variables
\begin{multline*}
\tilde{Q}(\theta;x,y) = \sum_{i \in \mathcal{V}} \sum_{t=1}^K  \sum_{z} q(z) \mathbf{1}(z_{i}=t) \log P(y_{i}|x_{i},z_{i}=t;\theta)
 -\lambda\sum_{(i,j)\in \mathcal{E}} \sum_{t=1}^{K}  \sum_{z} q(z) (b_{it}-b_{jt})^2  \mathbf{1}(z_{i} = z_{j} =t),
\end{multline*}
and then we get the marginal probabilities 
\begin{align*}
\tilde{Q}(\theta;x,y)& = \sum_{i \in \mathcal{V}} \sum_{t=1}^K q(z_i=t) \log P(y_{i}|x_{i},z_{i}=t,\theta) -\lambda\sum_{(i,j)\in \mathcal{E}} \sum_{t=1}^{K} (b_{it}-b_{jt})^2 q(z_i=z_j=t).
\end{align*}
Using \eqref{eq:p_y_x} and multiplying by minus equation $\tilde{Q}(\theta;x,y)$
we get the negative expected log-likelihood function in \eqref{eq:Q}.